\documentstyle[12pt]{article}
\begin{document}
\noindent
{\Large\bf Cut-off free finite zero-point vacuum energy and
the cosmological missing mass problem}\\~\\ {\Large N.
Kumar}\\ {\it Raman Research Institute, C V Raman Avenue,
Bangalore 560080, India}\\ {\it(nkumar@rri.ernet.in)}\\

\noindent
ABSTRACT\\
\noindent
As the mass-energy is universally self-gravitating, the
gravitational binding energy must be subtracted
self-consistently from its bare mass value so as to give
the physical gravitational mass. Such a self-consistent
gravitational self-energy correction can be made
non-perturbatively by the use of a gravitational `charging'
technique, where we calculate the incremental change $dm$
of the physical mass of the cosmological object, of size
$r_o$ due to the {\em accretion} of a bare mass $dM$,
corresponding to the gravitational coupling-in of the
successive zero-point vacuum modes, i.e., of the Casimir
energy, whose bare value $\Sigma_{\bf k}
\hbar ck$ is infinite. Integrating the `charging' equation, $dm
= dM - (3\alpha/5)Gm\Delta M/r_o c^2$, we get a
gravitational mass for the cosmological object that remains
finite even in the limit of the infinite zero-point vacuum
energy, i.e., without any ultraviolet cut-off imposed.
Here $\alpha$ is a geometrical factor of order unity.
Also, setting $r_o = c/H$, the Hubble length, we get the
corresponding cosmological density parameter $\Omega \simeq
1$, without any adjustable parameter. The cosmological
significance of this finite and unique contribution of the
otherwise infinite zero-point vacuum energy to the density
parameter can hardly be overstated.
\vspace{0.4cm}

\noindent
1~~~~INTRODUCTION\\
The single most important quantity determining cosmological
evolution is the average mass density ($\rho$) of the
universe, usually re-expressed as the dimensionless density
parameter $\Omega = \rho/\rho_c$, where $\rho_c \equiv
3H^2/8\pi G$ is the critical mass density required to
spatially close the universe. Here H is the Hubble {\em
constant}, the other observationally most important
expansion parameter.  Thus, with $H = 100 h km^{-1} s^{-1}
Mpc^{-1}$, we have $\rho_c$ = 2.10 x 10$^{-29} h^2g$
cm$^{-3}$, with $h$ of order unity.  It has been known for
sometime now that  the visible baryonic matter contents of
a typical spiral galaxy hardly add up to about one tenth of
the actual mass of the galaxy, estimated dynamically from
the rotation curve of its outer spiral arms. Other
cosmological considerations suggest that the mass of 
galaxies, including that of the invisible matter, in which
the galaxy seems embedded, in turn constitutes a mere one
fifth of the closure mass density of the universe.  Thus
most of the gravitational mass of the universe is missing.
This missing-mass problem has forced cosmologists to
propose the existence of a variety of non-baryonic dark
matter (see Sciama, 1993) $-$ the Cold Dark Matter (CDM),
the Hot Dark Matter (HDM) along with various candidate
particles, e.g., neutrinos, axions and other exotic extended
objects such as cosmic domain walls and cosmic strings, that
bear on structure formation.  Recent developments in
cosmology involving in particular the high red-shift
observations at accurately known cosmological distances,
made possible by the discovery of Type Ia supernovae
serving as {\em standard candles}, strongly favour $h$ =
0.65, and, most importantly, a density parameter value
$\Omega$= 1 (for a critical discussion on this, see Coles,
1998). Such a critical choice is also indicated from
considerations of the inflationary Big Bang models, and not
a little from the criteria of aesthetics and naturalness.
Normally $\Omega$ = 1 would  imply a spatially flat, ever
expanding and ever decelerating Friedmann universe. But
recent observations also suggest a possibly accelerated
expansion (see Coles, 1999). This has forced cosmologists
to re-introduce Einstein's cosmological constant $\Lambda$,
the third most important parameter now (See Peebles, 1999).
This has been invoked as an {\em X-matter}, and
observations support $\Omega = \Omega _\Lambda + \Omega_m$
=  1 with $\Omega_\Lambda$ = 0.7, and $\Omega_m$ = 0.3, the
remaining mass density including dark matter.  Now, the
cosmological constant is quintessentially a vacuum energy,
accumulated from the successive first-order phase
transitions that the inflationary universe is believed to
have undergone following the hot Big Bang. The high energy
calculations for this quintessential matter density $\big
(\Omega_\Lambda = \frac{\Lambda}{3H^2\big )}$, however,
give a value much too large, a factor $\sim 10^{120}$ times
greater than the total density of all the matter in the
universe. Thus, $\Omega, H$ and $\Lambda$ continue to
remain  three free parameters of modern cosmology $-$
getting continually constrained by observations, but
uncertainty prevails still. Be that as it may, it is a
compelling thought among cosmologists now that the universe
does seem, after all, pervaded by some kind of vacuum
energy that accounts for the missing mass. In this note we
address this problem, and examine the possibility of this
vacuum energy being none other than the familiar zero-point
energy of the physical vacuum associated with some
quantized Bosonic field, e.g., the photonic vacuum, and
show that its otherwise infinite and, therefore, worrisome
value gets renormalized self-gravitationally down, to yield
a cosmological density parameter $\Omega$ which is not only
finite, but in fact close to unity, and essentially free of
any adjustable parameter. We are talking here of a
cosmological Casimir Effect resolving the missing mass
problem!

A cosmological role for the zero-point energy has been
considered in the past (Kumar, 1969a, 1969b;  see also 
Mostenpanenko and Trunov, 1988; and Milonni, 1994). In
order to fully appreciate the point of this proposal, let
us recall that the bare zero-point energy of the
electromagnetic vacuum $E_{ZP} = \Sigma_{\bf k} \hbar ck$
is obviously infinite.  The cut-off wavelength $\lambda_c =
2\pi/k_c$ required to give an $\Omega \simeq$ 1 is
$\lambda_c \simeq$ 0.2 mm which is astrophysically
completely unacceptable on grounds of local Lorentz
invariance (Wesson, 1991). A cut-off of $\lambda_c \sim
\ell_{planck} \equiv (\hbar G/c^3)^{1/2}$ while acceptable
astrophysically would give a density parameter which is
practically infinite. While this infinity can be, and is,
subtracted away through a re-definition of the zero of
energy, or equivalently by normal ordering as is usually
done in quantum-field theory, no such subtraction is
permissible here as the non-linearly self-coupled
gravitational effect must be absolute, and it is large. But
the effect is, however, conspicuous by its absence here.
Indeed, the very reality of the zero-point energy has been
doubted for this very reason (see Milonni, 1994. But, the
zero-point  energy is fundamental, arising due to the
non-commutativity of the electric and the magnetic field
operators (in principle much the same way as the zero-point
energy of the condensed matter,e.g., of the quantum liquid
$^4$He, is due to the non-commutativity of the atomic
position and the corresponding momentum operators. Of
course, in case of the condensed matter it is possible to
directly measure its contributions to the cohesive energy
inasmuch as the condensed phase can be assembled from the
dispersed gas phase).  Then there is the well known
laboratory evidence for its local effect $-$ the Casimir
force between plates of a capacitor due to the change of
boundary conditions and the associated vacuum-mode
{\em depletion}.  It is, therefore, most compelling to accept the
reality of the zero-point energy.  The problem really is
with its being infinite!  It has been suggested that this
infinity can be cancelled exactly by invoking
supersymmetry, but the latter is clearly badly broken in
nature. The only way out now is to show that this {\em
bare-ly} infinite zero-point energy gets gravitationally
self-renormalized to a finite value which is acceptable
cosmologically. This indeed turns out to be the case as we
will show now.
\vspace{0.4cm}

\noindent
2~~~~DERIVATION\\
Consider the cosmological object (the universe) of size
$r_o$ having initially a gravitational mass $m = m_o$ (of
presumably baryonic origin) in the absence of any
gravitational coupling to the zero-point modes of the
physical vacuum.  We will now  follow the gravitational
{\em charging} of this object by gravitationally coupling
in the zero-point modes incrementally. Let at some stage of
the charging the current mass $m$ of the object be
incremented by $\Delta m$ due to the coupling in, or {\em
accretion}, $\Delta M$ from the vacuum modes. We have then
the {\em charging} equation
\begin{equation}
\Delta m = \Delta M - (6\alpha/5) Gm \Delta M/r_oc^2.
\end{equation}
It is essential here to realize that this {\em charging}
procedure is for a fixed $r_o$.  Here $\alpha$ is a
geometrical factor of O(1), its Euclidean value being 1.
The equation (1) can be readily integrated to give a
renormalized (gravitationally self-energy corrected) mass
$m$ for the object:
\begin{equation}
\ell n \frac{(1 - \frac{6\alpha}{5} Gm_o/r_o c^2)}{(1 -
\frac{6\alpha}{5} Gm/r_o c^2)} =
\frac{6\alpha}{5}\,\,\frac{GM}{r_oc^2}.
\end{equation}
>From Eq. (2) we readily see that as the {\em accreted}
gravitational {\em charge} $m$ increases from zero to the
full bare value of the zero-point energy (which is
infinite), the gravitational mass $m$ of the cosmological
object increases from $m_o$ to \(m(\infty) = \big
(\frac{5}{6\alpha}\big ) \frac{r_oc^2}{G}. \). This
renormalized value corresponds to a mass density \(\rho =
\big (\frac{5}{2\alpha \beta}\big ) \frac{c^2}{4\pi
r_o^2}\), where $\beta$ is another geometrical factor of
0(1), its Euclidean value being 1. Taking the size of the
object to be $r_o = c/H$ the Hubble length, we get for the
density parameter
\begin{equation}
\Omega = \frac{\rho}{\rho_c} = \frac{5}{3}\big
(\frac{1}{\alpha\beta}\big ).
\end{equation}
This is our main result.
\vspace{0.4cm}
  
\noindent
3~~~~DISCUSSION AND CONCLUSIONS\\ From Eq. (3), $\Omega$ =
5/3 for $\alpha = \beta$= 1, their Euclidean value.  Thus,
we get the renormalized cosmological mass density not only
finite, but also the corresponding value of the density
parameter $\Omega$ turns out to be close to unity!  It is
to be emphasized here that inasmuch as all zero-point
energies must gravitate universally, the above result is
quite independent of the particular quantized field, e.g.,
the photons here, with which the zero-point energy is
associated. Also, it should be remarked here that this
global (cosmological) effect in no way alters the local
(laboratory) Casimir effect that involves local changes of
the boundary conditions for the quantized modes, e.g., at
the metallic plates of the capacitor.

It is apt to contrast here our gravitational {\em charging}
procedure \underline{for the} \underline{cosmological
object} as in Eq. (1) with the one that underlies the ADM
mass (Arnowitt {\em et al}., 1960) of a self-gravitating
\underline{non-cosmological object} placed in an
asymptotically flat space, namely
\begin{equation}
dm = dM - (6/5) Gm dm/c^2r_o.
\end{equation}
Notice the subtle difference between Eqs. (1) and (4), in
that our Eq. (1) contains $\Delta M$, and not $\Delta m$,
in the second term on the RHS. Our charging procedure is
consistent with the physical idea that adding on a bare
mass dM to the pre-existing gravitational mass $m$ should
give a gravitational mass-defect proportional to $m dM$.
Equation (4) can be readily integrated to give the
well-known ADM mass $m$, that tends to $\infty$ as $M$
tends to infinity. This difference is remarkable and calls
for further examination.

Finally, it is to be noted that we have considered here
only the overall cosmological consequence of the zero-point
energy. Its association with the baryonic matter, important
for structure formation, has not been touched upon.

In conclusion we have shown that the zero-point energy of
the vacuum $-$ the Casimir energy $-$ is gravitationally
self-renormalized to yield a finite value for the density
parameter $\Omega$ which turns out to be close to unity,
without any adjustable parameter such as the ultraviolet
cut-off. This gravitational self-energy correction is
carried out self-consistently through a {\em charging}
procedure. This is a step forward towards resolution of the
missing mass (Dark Matter) problem in cosmology.  One could
perhaps say that the Dark Matter is after all made of Light
$-$ the zero-point photon energy!
\newpage
\noindent
{\bf REFERENCES}\\
\noindent
Arnowitt R., Deser S. and Misner C., 1960, Phys. Rev., 120,
321\\ 
Coles P., 1998, Nature, 393, 741\\ Kumar N., J.
Phys., A2 (1969a) 210\\ Kumar N., 1969a, Progr. Theor.
Phys., 41, 382\\ 
Milonni P.W., 1994, The Quantum Vacuum, An Introduction to
Quantum\\ \hspace*{0.4cm}Electrodynamics, Acad. Press, Inc., Boston\\
Mostenpanenko V.M. and Trunov N.N., Sov. Phys. Usp., 1988,
31, 965\\ 
Peebles P.J.E., 1969b, Nature, 398, 25\\
Sciama D.W., 1993, Modern Cosmology and the Dark
Matter Problem,\\ \hspace*{0.4cm}Cambridge University Press, Cambridge\\
Wesson P.S., 1991, Astrophys. J., 378, 466\\
\end{document}